\documentclass[aps,prd,twocolumn,superscriptaddress,altaffilletter,nobibnotes,nofootinbib,10pt,showpacs,showkeys,preprintnumbers]{revtex4-2}

\usepackage{float}
\usepackage{graphicx}% Include figure files
\usepackage{dcolumn}
\usepackage[dvipsnames]{xcolor}
\usepackage{bm}% bold math
\usepackage{lineno}
\usepackage{float}
\usepackage{amsmath}
\usepackage{soul}
\RequirePackage{multirow}

\usepackage{tikz,xcolor,hyperref}

\definecolor{lime}{HTML}{A6CE39}
\DeclareRobustCommand{\orcidicon}{
	\begin{tikzpicture}
	\draw[lime, fill=lime] (0,0) 
	circle [radius=0.16] 
	node[white] {{\fontfamily{qag}\selectfont \tiny ID}};
	\draw[white, fill=white] (-0.0625,0.095) 
	circle [radius=0.007];
	\end{tikzpicture}
	\hspace{-2mm}
}
\foreach \x in {A, ..., Z}{\expandafter\xdef\csname orcid\x\endcsname{\noexpand\href{https://orcid.org/\csname orcidauthor\x\endcsname}
			{\noexpand\orcidicon}}
}

\usepackage{soul}
\usepackage{float}

\definecolor{lime}{HTML}{A6CE39}
\DeclareRobustCommand{\orcidicon}{
	\begin{tikzpicture}
	\draw[lime, fill=lime] (0,0) 
	circle [radius=0.16] 
	node[white] {{\fontfamily{qag}\selectfont \tiny ID}};
	\draw[white, fill=white] (-0.0625,0.095) 
	circle [radius=0.007];
	\end{tikzpicture}
	\hspace{-2mm}
}
\foreach \x in {A, ..., Z}{\expandafter\xdef\csname orcid\x\endcsname{\noexpand\href{https://orcid.org/\csname orcidauthor\x\endcsname}
			{\noexpand\orcidicon}}
}

\begin{document}
%\preprint{APS/123-QED}

\title{Study of density independent scattering angle and energy loss  for low- to high-$Z$ material using Muon Tomography}

\author{Bharat Kumar Sirasva}
\affiliation{Department of Physical Sciences, Indian Institute of Science Education and Research (IISER) Mohali, Sector 81 SAS Nagar, Manauli PO 140306 Punjab, India}
\author{Satyajit Jena\orcidB{}}
\email{sjena@iisermohali.ac.in}
\affiliation{Department of Physical Sciences, Indian Institute of Science Education and Research (IISER) Mohali, Sector 81 SAS Nagar, Manauli PO 140306 Punjab, India}
\author{Rohit Gupta\orcidA{}}
%\email{ph15049@iisermohali.ac.in}

\affiliation{Shaheed Mangal Pandey Government Girls Degree College (SMPGGDC), Jananayak Chandrasekhar University (JNCU), Ballia 277001, Uttar Pradesh, India}

%\affiliation{%

% ----------- Write the Abstract Here -------------------------
\begin{abstract}
Cosmic ray muon, as they pass through a material, undergoes Multiple Coulomb Scattering (MCS). The analysis of muon scattering angle in a material provides us with an opportunity to study the characteristics of material and its internal 3D structure as the scattering angle depends on the atomic number, the density of the material, and the thickness of the medium at a given energy. We have used the GEANT4 toolkit to study the scattering angle and utilize this information to identify the material. We have analyzed the density dependent $\&$ density independent scattering angle and observed various patterns for distinct periods in the periodic table.
\end{abstract}
%\pacs{05.70.Ce, 25.75.Nq, 12.38.Mh}

\maketitle

\section{\label{sec:intro}INTRODUCTION}
Muon tomography is a novel technique developed to study the internal structure of a material \cite{Tomo, Tomo1}. With the help of this approach, we can independently estimate a material's scattering density  \cite{scat} and use that information to create a 3D image \cite{imag} of the material and accurately predict its characteristics. This technique employs cosmic ray muons, which are of interest because of their larger abundance on earth \cite{crm} and higher  penetration capability. Cosmic ray bring large flux of muons to our earth's surface \cite{Kampert:2000tw}. As the muon pass through a material it undergoes Multiple Coulomb Scattering(MCS) \cite{MCS} and this serves as a backbone for muon tomography. Many detectors, such as scintillator detectors \cite{scintillator}, MRPC detectors \cite{mrpc}, and gas detectors \cite{Gas, Gas1}, are utilised for muon tomography investigations to determine material type. The essential ingredients for the tomographic configurations are muon scattering with materials, muon scattering angle, and muon kinetic energy. As the muon pass through a material, it interact with the constitutes atoms molecules and transmit energy. The estimation of energy loss by muon as it passes through a material is based on the total amount of energy lost in each contact between muon \& atom and is governed by the Bethe - Block eq. \cite{Bethe}. 

In the current study, we have estimated the scattering angle for muons in various materials. As the muon passes through material, its trajectory will be perturbed by nucleus. Hence the scattering angle and scattering density is different for different material \cite{density}. Muon paths are significantly deflected due to multiple Coulomb scattering in the material. In order to determine the muon's deflection angles and energy loss, we run Monte Carlo simulations using the GEANT4 \cite{Geant4} toolkit package, and we follow a technique that can be replicated in an experiment depending on where the muons hit the detector layer. Our primary goal is to study the muon's energy absorption in material and their path deviation by multiple coulomb scattering. For this purpose, we first setup our physics list $FTFP\_BERT$ which handles all the interaction in GEANT4 simulation. Using the simulating package and libraries, we next create an experimental virtual setup. And then we allow muon to pass through the detector and study the scattering angle. The data on scattering angle and energy deposition in this study are analysed using ROOT  \cite{Root}, which is a data analysis framework developed at CERN.

The following is an outline of this study: section 2 introduces the theory of multiple Coulomb scattering, A brief introduction about the GEANT4 simulation and the fundamental aspects of our simulation setup is provided in section 3. The result of the study is provided in section 4 followed by the conclusion in section 5.

\section{Theory}
When particles travelling through matter are subject to repeated elastic Coulomb scattering from the nuclei, each scattering centre produces a little deviation in the approaching particle's trajectory. There are three possible scenarios for scattering: First, a single scattering can be considered in cases where the scattering is really tiny and the likelihood of many interactions is extremely low. The next scenario involves a case where the amount of scattering rises but stays below a few tens of contacts, which is referred to as plural scattering \cite{plural}. And the last is multiple coulomb scattering, when net energy loss remains low, but the path deviation is significant because of the muon's multiple interactions with material. 

It is presumed that the scattering angle will affect the particle's travel length through the material. As the particle passes through various material, its scattering characteristic and the energy spectrum can be employed in imaging to discriminate between various materials and densities \cite{Clarkson:2013mta}. We primarily use these characteristics in our research on muon scattering. As the muons interact with materials, they cause both energy loss and simultaneous scattering by the nuclei and electrons present in the materials. In cargo inspections, muon interactions with other materials are predominantly controlled by Coulomb scattering. So, with each scattering, the muon track will be slightly diverted from its initial course. As a result, the deflection angle between the incoming and outgoing directions is detectable during repeated Coulomb scattering, which may be employed in small-scale object imaging \cite{RSI}. The mathematical formulation for scattering angle comes from Moliere theory \cite{lynch}. According to this theory, the scattering angle approximates the Gaussian distribution with mean zero. And the standard deviation of the scattering angle is given by Moliere equation \cite{Tripathy:2016gpw,sehgal}.
 
\begin{equation} \label{eq:1}
\centering
    \theta_0 = \frac{13.6 MeV}{\beta pc} \sqrt{\frac{L}{X_0}}\left[1 + 0.038\ln\left(\frac{L}{X_0} \right) \right]
\end{equation}
where $p$ is momentum of the muon, $L$ is thickness of the material, and $X_0$ is the radiation length of the material and is given by Eq. \ref{eq:2}.
\begin{equation} \label{eq:2}
\centering
    X_0 = \frac{A\times 716.4g/cm^2}{\rho.Z(Z+1)ln(287/\sqrt{Z})}
\end{equation}
 Where $A$ is the mass number, $Z$ is the atomic number and $\rho$ is the density. The energy loss and multiple coulomb scattering of muons as it passes through a given detector geometry can be simulate in GEANT4. A brief introduction of GEANT4 simulation and the detail of detector geometry used for current analysis is provide in next section.

\section{Geant4 Simulation}
Geant4 is an object-oriented simulation toolkit for detector physics and high-energy physics \cite{Geant4}. Various aspects of detector physics are considered when defining a simulation process in the Geant4 toolkit. This includes geometry of system, the material involved, the fundamental particle of interest, the response of sensitive detector components, the generation of event data, the storage of events and tracks, and the visualisation of detector and particle trajectories. The advantage of Geant4 is that it can handle particle-matter interactions over a large range. The Geant4 toolkit is a flexible and comprehensive software package designed for simulation applications involving particle interaction and transition through materials. It can handle complicated geometries efficiently and compactly, and it supports the visualisation of geometry and particle tracks via a number of interfaces. In Geant4, there is a category for the modelling of particle-matter interaction that covers the system of units, constants, physical characteristics of particles, and materials. Physical interactions such as electromagnetic interactions between leptons, photons, hadrons, and ions, are addressed by the category process. Geant4 also handles the pile-up problem.
 \begin{figure}[!h]\label{fig:1}
 \centering
  \includegraphics[width=.6\linewidth]{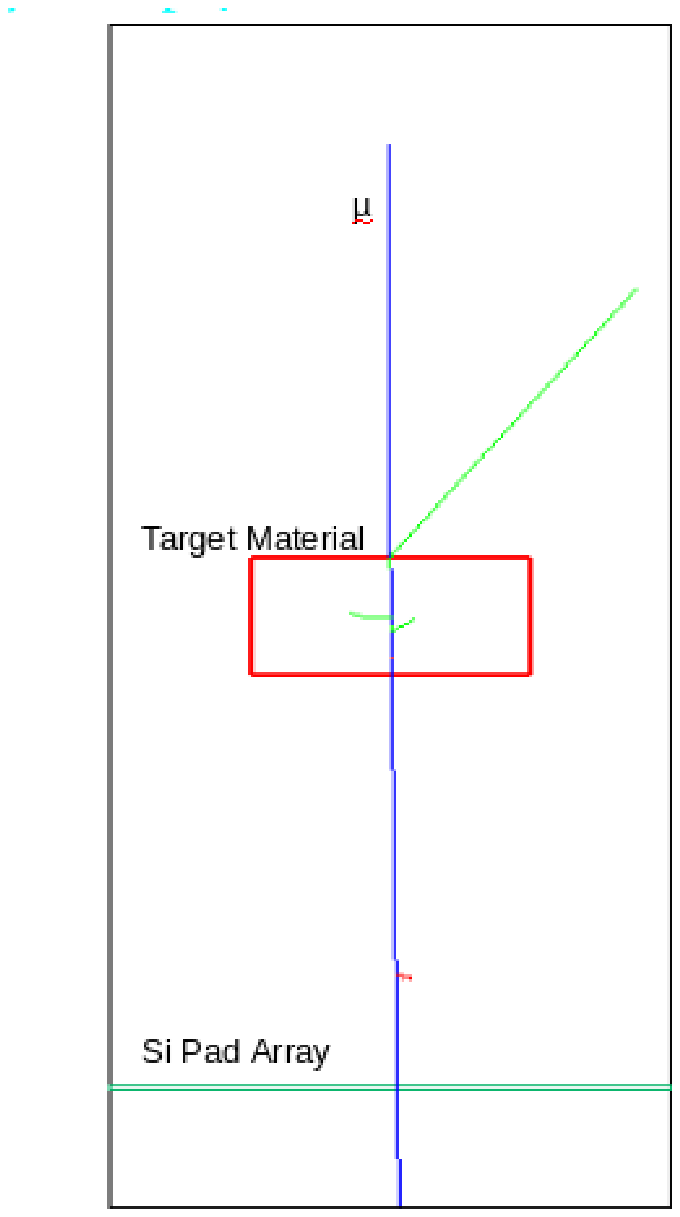}
  \caption{Detector geometry defined in Geant4 }
    \label{fig:my_label}
\end{figure}

As discussed in the privious section, the simulation performed as a part of the present analysis is based on muon's multiple coulomb scattering(MCS) The default multiple scattering model in the Geant4 simulation tool is based on the G4MultipleScattering class. This model simulates particle scattering after a particular step and computes the mean length correction and mean lateral displacement.

The detector geometry used for this study is presented in Fig. 1. The detector setup includes a muon generator, a target material and a silicon pad array. The detector setup is placed in the "World" volume. The world is represented by the outer white box, which we define as "G4 air", and muons are created by the G4Particle gun in the world volume. The G4Particle Gun is placed 20 cm away from the middle of the material. The blue line depicts the muon track. The momentum direction of incoming muon is only in $y$ direction and energy is 5 $GeV$ for the incoming muons. The red box is for target material and it is added in Geant4 at the centre of geometry. The dimensions of the target material is \(10cm\times5cm\times10cm\). The horizontal green box is a silicon pad array used to measure the position of scattered muons. The width of the $Si$ pad array is \(2mm\) which is placed at 20 cm below from the material. The muon primarily interacts with the material, transfers energy via repeated Coulomb scattering, deviates from its original track and strikes the silicon pad array.

 In this section we discussed about the detector setup for our study and the analysis of the data produces after running the simulation for the above detector arrangement is discussed in the next section.
\section{Data Analysis And Result }
As the muon is allowed to pass through $Si$ pad array the Geant4 simulation gives us the kinematic quantities of interest such as the direction of the momentum of the incoming and scattered muons, the location of the muon impact at the silicon pad, and the energy loss by muon as it traverse through the target material. One can obtain scattering angle by using the momentum of incoming and scattered muon. Another method employed to determine the scattering angle is by using the detector geometry. In this method, we measure the deflection in trajectory of incoming muon as it passes through the detector and determine the scattering angle using trigonometric properties.

Considering $y$ as the axis representing the direction of muon propagation, its deflection will be along the $x$-$z$ plane. As the muon of energy 5 $GeV$ is allowed to pass through the target material, it scatters along the $x$-$z$ plane and the corresponding scattering angle $\theta_x$ and $\theta_z$ axis is given as:
\begin{equation}\label{eq:3}
    \theta_x = tan^{-1}\left(\frac{x}{y}\right) \quad\text{and}\quad    \theta_z = tan^{-1}\left(\frac{z}{y}\right) 
\end{equation}
Here, $x$, $z$ are the scattered muon locations at the $Si$ pad array, whereas $y$ is the distance between the material and the $Si$ pad array. Fig. 2(a) and 2(b) presents the distribution of muon scattering angle along $x$ and $z$ direction respectively. From Fig. 2 we observe that the mean deflection angle is close to zero for both direction however the standard deviation increases for heavier material.

\begin{figure}[!h]
    \centering
    \includegraphics[scale = .42]{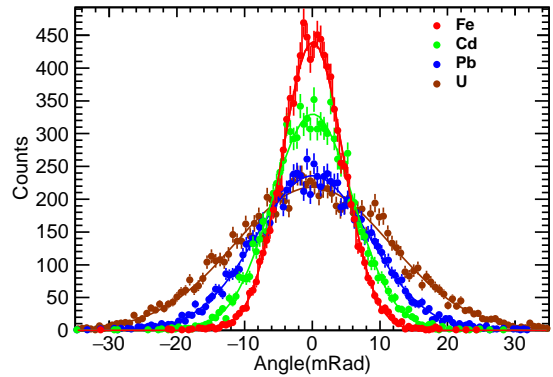}
    
    (a)
    
     \includegraphics[scale = .42]{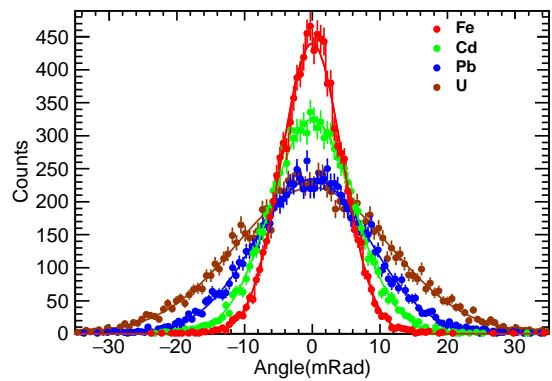}
     
     (b)
    \caption{Distribution of muon scattering angles in the $x$ (Fig. 2(a)) and $z$ (Fig. 2(b)) direction for the materials $Fe$, $Pd$, $Pb$, and $U$. Although the standard deviation increases with heavier material, the mean of this Gaussian distribution is quite near to zero.}
    \label{fig:my_label}
\end{figure}
\begin{table}[!h]
\caption{Data on the angle of muon scattering in the x direction. }
\centering
%\begin{ruledtabular}
\begin{tabular}{ccc}
% &\multicolumn{2}{c}{$D_{4h}^1$}&\multicolumn{2}{c}{$D_{4h}^5$}\\
 Material &  Mean (mRad)& Sigma (mRad)\\ \hline
 Fe&0.0821$\pm$ 0.0452  &4.5032$\pm$ 0.0346  \\
 Cd &0.0802$\pm$ 0.0603    & 5.9807$\pm$ 0.0445  \\
 Pb & 0.0493$\pm$ 0.0845  & 8.3736$\pm$ 0.0630  \\
 U & 0.0667$\pm$ 0.1158 & 11.4556$\pm$0.0814  \\
 
 %Boltzmann&$\pi^-$ $(0$ to $5\%)$&$245.873$&$-$&$-$&$-$&$-$&$-$&$24.4878$\\
\end{tabular}
%\end{ruledtabular}

\end{table}

\begin{table}[!h]
\caption{Data on the angle of muon scattering in the z direction.}
\centering
%\begin{ruledtabular}
\begin{tabular}{ccc}
% &\multicolumn{2}{c}{$D_{4h}^1$}&\multicolumn{2}{c}{$D_{4h}^5$}\\
 Material &  Mean (mRad)& Sigma (mRad)\\ \hline
 Fe&0.0271$\pm$ 0.0451 & 4.4901$\pm$ 0.0338 \\
 Cd & 0.0081$\pm$ 0.0612 & 6.0891$\pm$ 0.0461 \\
 Pb & 0.0465$\pm$ 0.0849 & 8.4141$\pm$0.0632 \\
 U & 0.0540$\pm$ 0.1142 & 11.3152$\pm$0.0842 \\
 
 %Boltzmann&$\pi^-$ $(0$ to $5\%)$&$245.873$&$-$&$-$&$-$&$-$&$-$&$24.4878$\\
\end{tabular}
%\end{ruledtabular}

\end{table}
Instead of finding the components $\theta_x$ and $\theta_y$, we can also determine the overall scattering angle $\theta$ by calculating $r$. As already discussed, we can determine $\theta$ either by using the momentum of incoming and outgoing muon (Eq. \ref{eq:4}) or by using the geometry of hit point on $Si$ pad array (Eq. \ref{eq:5}). 
\begin{equation}\label{eq:4}
       \theta = cos^{-1}\left(\frac{p_1.p_2}{|p_1| | p_2 |} \right)
\end{equation}

\begin{equation}\label{eq:5}
    r = \sqrt{x^2 + z^2} \quad\text{and}\quad  \theta = tan^{-1}\left(\frac{r}{y}\right)
\end{equation}

In the above  equation, $p_1$ and $p_2$ represent the momentum of incoming and scattered muons, respectively. The angle distribution of muon for different materials estimated using Eq. \ref{eq:5}  is presented in Fig. 3 and we observe the heavier material has a larger deflection angle as well as larger standard deviation. Further, the distribution for $U$ is wider than that of lighter nuclei $Fe$. This indicates that for heavier nuclei, the probability of muon scattering is higher, as a result the most probable value of scattering angle and standard deviation is higher for heavier material. Also, the number of entries at the most probable scattering angle value is fewer in heavier nuclei compared to the lighter nuclei because of the wider spectra in heavier nuclei. Our outcome using Eq. \ref{eq:5} is comparable to the result obtained in the study of waste nuclear barrels with heavier material, which computed the scattering angle using Eq. \ref{eq:4}, as shown in Ref. \cite{Topuz:2021ypu}.  

\begin{figure}[h]
    \centering
    \includegraphics[scale = .42]{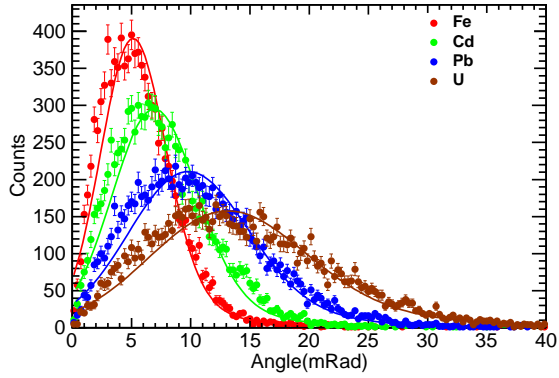}
    \caption{Eqs. 4 and 5 provide the same outcome for muon angular distribution in $Fe$, $Pd$, $Pb$, and $U$ materials, resulting in this distribution. For uranium, the distribution broaden and iron gives a sharp peak.}
    \label{fig:my_label}
\end{figure}

\begin{table}[!h]
\caption{Muon Scattering angle data acquired from Figure 4.} 
\centering
%\begin{ruledtabular}
\begin{tabular}{ccc}
% &\multicolumn{2}{c}{$D_{4h}^1$}&\multicolumn{2}{c}{$D_{4h}^5$}\\
 Material &  MPV (mRad)& GSigma (mRad)\\ \hline
 Fe&4.5720$\pm$0.0519&2.6051$\pm$0.0427 \\
 Cd &6.1979 $\pm$0.0720 & 3.4115$\pm$0.0599 \\
 Pb & 8.7726$\pm$0.1007 & 4.8236$\pm$0.0771 \\
 U & 11.4959$\pm$0.1618 & 6.1497$\pm$0.1286 \\
 
 %Boltzmann&$\pi^-$ $(0$ to $5\%)$&$245.873$&$-$&$-$&$-$&$-$&$-$&$24.4878$\\
\end{tabular}
%\end{ruledtabular}

\end{table}

From Fig. 2, we observe that the scattering angle along $x$ and $z$ direction looks like a Gaussian distribution and the fit using the Gaussian function in ROOT shows good agreement between the data and the fit function. Considering the transformation to polar coordinate, we get angle $\theta$ instead of components $\theta_x$ and $\theta_z$. While studying the component of scattering angle \(\theta_x \) \& \(\theta_z\) we only have one measurable quantity, the standard deviation, the mean is approximately zero for all materials, see Table 1 \& 2. Whereas in case of Fig. 3 we get two measurable parameter, the most probable value of scattering angle \(\theta\) and  the standard deviation. A convoluted function of Landau and Gaussian distributions (Langauss function available in ROOT) is used to fit the data. In practise, the convoluted Landau and Gaussian fitting functions are used to evaluate the distribution and provide a good fit to the asymmetric distribution.The most probable (MP) value and the Gsigma value are the two parameters we obtain from the fitting. 10000 events are used to calculate the parameter Most Likely value and the Gsigma value of the Langauss distribution. It is advantageous for the analysis of material identification to have two quantifiable parameters rather than only one. 

\begin{figure}[!h]
 \centering
  
    \includegraphics[scale = .42]{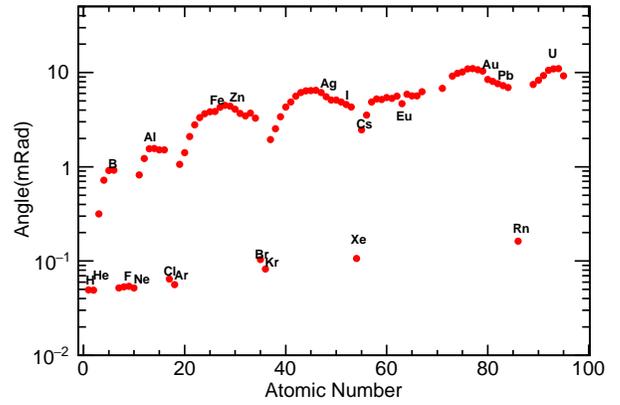}
    (a)
    \includegraphics[scale = .42]{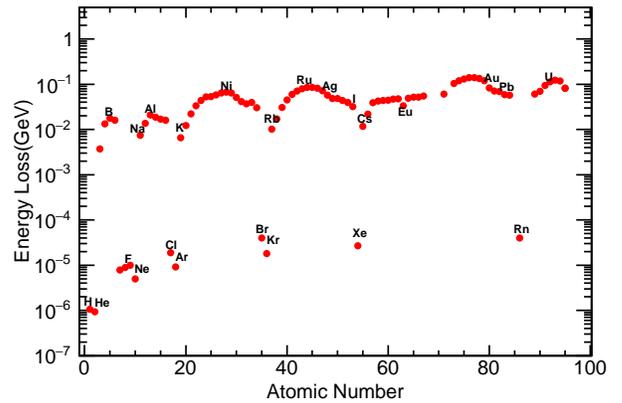}
    (b)
 \caption{The variation of scattering angle of muons with atomic number (Fig. 4(a)), and the muon energy loss versus atomic number (Fig. 4(b)).}
    \label{fig:my_label}
\end{figure}
We have also studied the variation of muon scattering angles with the atomic number for a broad range of materials. From Fig. 4(a), we observe that the plot of scattering angle versus atomic number shows various peaks and valleys. There are several factors affecting the scattering angle, such as the type of material, the size of the atoms, and the strength of the bonding between the atoms. If we classify elements into groups such as halogens, rare earth elements, noble gases, and transition materials etc, we observe that for a particular group the scattering angle increases with increase in atomic number.  We have also studied the variation of muon energy loss with atomic number and the result is presented in Fig. 4(b). The energy loss is very small compared to the muon's original energy. For H, the energy loss is approximately \(10^{-6} GeV\) for muons with an initial energy of 5 $GeV$. The energy loss increases for heavier material with $Fe$, $Cd$, $Pb$ and $U$ having energy loss 0.053 $GeV$, 0.057 $GeV$, 0.069 $GeV$ and 0.133 $GeV$ respectively.

\begin{figure}[!h]
 \centering
  \includegraphics[scale=.42]{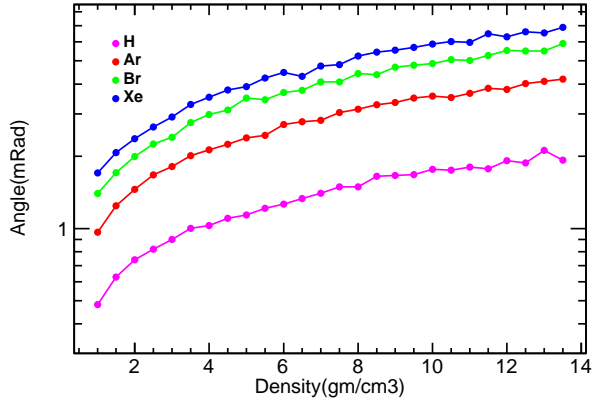}
  (a)
  \includegraphics[scale = .42]{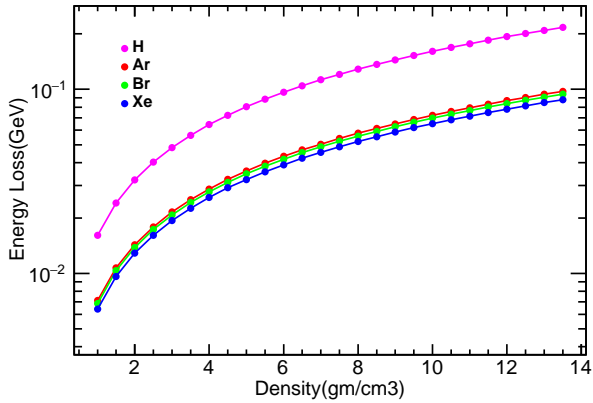}
    (b)
     \caption{The Fig. 5(a) shows the variation of the scattering angle of muons as a function of density and Fig. 5(b) shows the variation of energy loss as a function of density.}
    \label{fig:my_label}
\end{figure}

  We have also analyzed the density dependency of energy loss and the scattering angle for different materials. The density dependence for $H$, $Ar$, $Br$ and $Xe$ is shown in Fig. 5 and we observe that both the scattering angle and energy loss increases with rise in density of materials. At a density of 2 $g/cm^3$ the energy loss for $H$, $Ar$, $Br$ and $Xe$ is 0.032 $GeV$, 0.014 $GeV$, 0.013 $GeV$, and 0.012 $GeV$ respectively, and the scattering angle is 0.739 $mRad$, 1.454 $mRad$, 1.995 $mRad$, and 2.368 $mRad$. As density increases to $12 g/cm^3$, the energy loss increases to 0.193 $GeV$, 0.086 $GeV$, 0.083 $GeV$, and 0.078 $GeV$ and the scattering angle increases to 1.914 $mRad$, 3.803 $mRad$, 5.520 $mRad$, and 6.298 $mRad$ for $H$, $Ar$, $Br$ and $Xe$ respectively. From Fig. 5(a), we can see that the scattering angle is maximum for $Xe$ which is the heaviest of four elements.

The density-independent muon energy loss in materials is shown by our result in Fig. 6(b). Here, the energy loss in all materials is divided by their respective densities. From Fig. 6(b), we observe that the energy loss per unit density is the same for all materials and that the energy loss is substantially smaller for gases compared to solids.

\begin{figure}[!h]
 \centering
  \includegraphics[scale = .42]{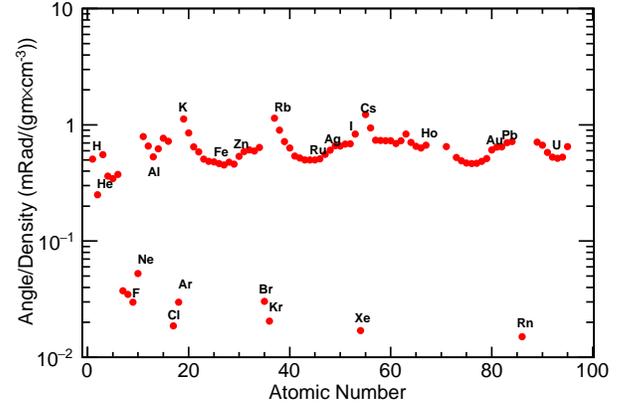}
   (a)
   \includegraphics[scale = .42]{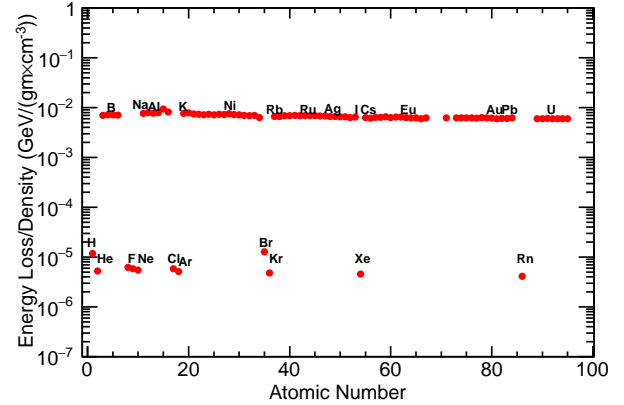}
   (b)
   \caption{The density-independent relationship between muon scattering angle and atomic number (Fig. 6(a)) and energy loss with atomic number (Fig. 6(b))}
\end{figure}

Since the scattering angle depends on the atomic size, the total number of atoms present, and the characteristics of the material, so the scattering angle per unit density will not be same for different material.
When we look at the density-independent muon scattering angle with the atomic number in Fig. 6(a), we can see that it changes with the atomic number and that there is a pattern.
The scattering angle is grater for alkali metals like $K$, $Rb$, and $Cs$, then it decreases for transition elements, slightly increases for p-block elements, and finally decreases for halogens and noble gases. The transition element has a valley in the fourth period of the periodic table, and the lowest value of density-independent scattering angle is for the sixth, seventh, and eighth groups in the periodic table. The scattering angle increases until $Se$, then drops for $Br$ and $Kr$. The same pattern can be seen for the fifth period in the periodic table. If we concentrate on the halogens $F$, $Cl$, $Br$, and $I$ in Fig. 6(a), the scattering angle for $F$ is greater than that of $Cl$, even though $Cl$ has a higher atomic number than $F$ and then for $Br$ the scattering angle increases because at room temperature $Br$ is in the liquid state, so the probability of interaction of a muon increases, and if we see $I$, it is in the solid state at room temperature, so the interaction is greater than the rest of the halogens. When noble gases are studied, the density-independent scattering angle decreases as the atomic number increases. Noble gases differ from other materials in the sense that they do not interact with each other. When we compare $Rn$ with $Ne$, we can observe that $Rn$ has a bigger atomic size, which implies it has fewer atoms in the same volume as $Ne$. As a result, muon scattering counts in Rn decreased. The scattering angle of a muon decreases as its scattering in Rn decreases. 
%The interaction of nearby atoms causes the scattering angle to rise for all other halogens, excluding Cl. 

%For heavier materials, fewer particles contact, but the energy is lost more quickly than for lighter materials.  
\begin{figure}[!h]
 \centering
  \includegraphics[scale = .42]{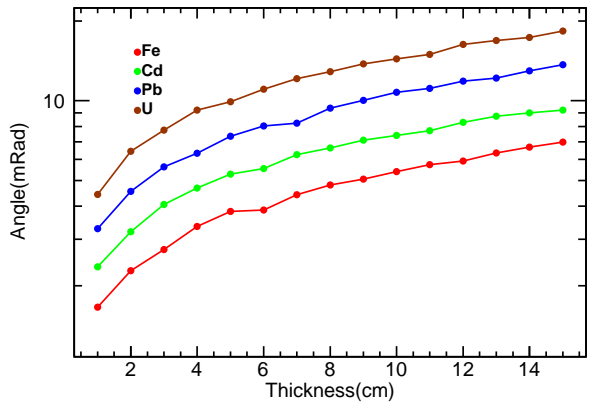}
  (a)
  \includegraphics[scale = .42]{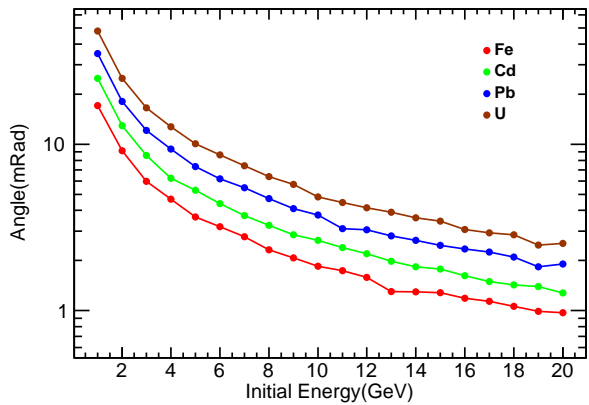}
   (b)
\caption{The upper plot shows how the scattering angle varies with material thickness. In the lower plot, the scattering angle of muons varies with their initial energy. Here, $U$ is a heavier material than $Pb$ and $Fe$, and as a result, the scattering angle is higher for $U$ in both cases. }
\end{figure}

We have also studied the variation of scattering angle with the thickness of material \& the initial energy of muons and the result are presented in Fig. 7(a) \& 7(b) respectively. We analyse $Fe$, $Cd$, $Pb$ and $U$ for thickness variation and we see, material's interaction with the muon increases with thickness because more atoms are included in a given volume as thickness increases. As a result, the scattering angle increase. In Fig. 7(b), the interaction of muon's, however, decreased when the thickness was fixed while the muon's initial energy was increased because the higher muon's initial energy resulted in a smaller scattering angle.

\section{Conclusion}
In this paper, we use our detector geometry and the scattering muon impact on the detector to simulate muon scattering angles using different methodologies. We examine the scattering angle of the muon with various materials and compute its energy losses in the material. We have also studied the variation of scattering angle different type of material thickness. Some of the important findings of our study are listed below.
\begin{itemize}
    \item We observed that the scattering angle component's distribution along the x and z axes follows a Gaussian distribution; however, when polar coordinates are taken into account, the distribution approaches a Landau distribution. We calculated the fit parameter by fitting $\theta_x$ $\&$ $\theta_z$ with the Gaussian distribution and for $\theta$ we use Langauss function. For the goal of material identification, it is preferable to have more quantifiable parameters, hence in practical applications like cargo scanning, it is preferable to use $\theta$ (see Fig. 3) when performing muon tomography. 
    \item The analysis of the scattering angle and energy loss density dependence reveals that both variables increase with material density, which is in accordance with the theory that as density rises, interactions with the material increase because there are more atoms in the same volume. As a result, the scattering angle and energy loss of muons increase with density. 
    \item The density-independent examination of energy loss and scattering angle with atomic number yields some interesting results. For all solids, the density-normalized energy loss is constant. The density-normalised scattering angle, on the other hand, exhibits patterns for distinct periods in the periodic table. Because of the lower bonding strength between gas atoms, the density-normalized scattering angle for gases falls with atomic size. Hence we can say that the chemical characteristics of the material play a crucial role in scattering angles. 
\end{itemize}

 \section{Acknowledgement}

  B. K. Sirasva would like to acknowledge Dr. Harpreet singh kainth for the fruitful discussions. B. K. Sirasva also like to acknowledge the financial support provided  by  UGC  through  fellowship no. 191620096154

% ****** End of file apssamp.tex ******
%\input{ref.tex}

\end{document}